\def\year{2021}\relax
\documentclass[letterpaper]{article} 
\usepackage{aaai21}  
\usepackage{times}  
\usepackage{helvet} 
\usepackage{courier}  
\usepackage[hyphens]{url}  
\usepackage{graphicx} 
\usepackage{natbib}  
\usepackage{caption} 

\usepackage{color}
\usepackage[super]{nth}
\usepackage{microtype}
\usepackage{color}
\usepackage{subfigure}
\usepackage{tabularx}
\usepackage{verbatim}
\usepackage{array,multirow,graphicx}
\usepackage{multicol}
\usepackage[page]{appendix}

\title{More than Meets the Tie: Examining the Role of\\Interpersonal Relationships in Social Networks}
\author {
    Minje Choi,
    Ceren Budak,
    Daniel M. Romero,
    David Jurgens \\
}
\affiliations {
    University of Michigan \\
    \{minje,cbudak,drom,jurgens\}@umich.edu
}
\begin{document}

\maketitle

\begin{abstract}
Topics in conversations depend in part on the type of interpersonal relationship between speakers, such as friendship, kinship, or romance.
Identifying these relationships can provide a rich description of how individuals communicate and reveal how relationships influence the way people share information.
Using a dataset of more than 9.6M dyads of Twitter users, we show how relationship types influence language use, topic diversity, communication frequencies, and diurnal patterns of conversations.
These differences can be used to predict the relationship between two users, with the best predictive model achieving a macro F1 score of 0.70.
We also demonstrate how relationship types influence communication dynamics through the task of predicting future retweets.
Adding relationships as a feature to a strong baseline model increases the F1 and recall by 1\% and 2\%.
The results of this study suggest relationship types have the potential to provide new insights into how communication and information diffusion occur in social networks.
\end{abstract}

\section{Introduction}
Dyadic relationship between individuals are a fundamental characteristic in online social networks such as Twitter.
For relationships, concepts such as tie strength~\citep{granovetter1973strength}, signs~\citep{leskovec10positivelinks,leskovec10signed}, and direction~\citep{foster10edgedirection} have been used to study communicative behaviors such as reciprocity~\citep{cheng2011reciprocity}, topic diffusion~\citep{romero13topical}, and echo chambers~\citep{colleoni14echochamber}. Individuals in these networks are largely organized around social structures such as work, neighborhood or families~\citep{feld81focisocialties,feld82socialstructural}, forming \textit{interpersonal relationships}, such as friendships, kinship, and romantic partnerships.
These interpersonal relationship types can influence communication and behavior in the network---e.g., consider what information might be shared between friends versus with a parent.
Knowing and inferring relationship types in a social network can have several implications, such as directing messages to the appropriate social audience
~\citep{Ranganath15socialfoci}, improving  information diffusion models, and detecting social communities ~\citep{tang12tiestrength}.
However, due to lack of data availability, interpersonal relationships have rarely been considered for these tasks in social network research.

In this paper, we aim to close this gap by inferring interpersonal relationship types from dyadic interactions in online social networks.
Indeed, several studies have tried to classify relationships in domains such as phone call logs~\citep{min13miningsmartphonedata}, chatroom conversations~\citep{tuulos04chatdatamining}, and conversation transcripts across messaging platforms~\citep{Welch2019LookWT}. 
While these studies show predicting relationships is possible for private exchanges or niche topic-based communities, general social media lead to additional challenges due to the substantially higher diversity in content and relationship types.
Also, while there has been prior work directly aiming to predict relationships from Twitter data~\citep{adali12relationships}, the predicted categories are data-driven clusters that do not directly correspond to known social relationship types.
Yet, as we will show, interpersonal communication still contains linguistic signals that reveal social relationships, enabling accurate prediction. 

The contributions of this study are as follows.
First, using a massive dataset of interactions between 9.6 million Twitter user dyads with labeled relationships, we conduct an extensive analysis of linguistic, topical, network and diurnal characteristics across relationship categories.
We show that relationships on Twitter follow existing theories of interpersonal relationships and reveal complex social dynamics.
Second, we introduce a neural network model for classifying five relationship types from linguistic and network features, achieving an F1 of 0.70, that substantially improves upon a strong classifier baseline (0.55) and random guess (0.20).
Finally, we show that knowing the type of relationship improves performance on the challenging task of predicting whether one user will retweet another's message, improving the F1 by 1.4\% for tweets that do not contain URLs and 2.0\% for tweets that do, and highlighting the benefit of modeling the interaction between relationship and content.
A pretrained version of our model is publicly available\footnote{https://github.com/minjechoi/relationships}.

\section{Interpersonal Relationships}
\label{sec:relationships}
Interpersonal relationships between Twitter users can be broadly grouped into five categories: \textit{social}, \textit{romance}, \textit{family}, \textit{organizational} and \textit{parasocial}.
These categories, based on prior theory from communication studies and sociology, cover the social relationships studied in both offline~\citep{knapp80communicationrelationships,feld82socialstructural} and online settings~\citep{ozenc11lifemodes}.
Examples for each category are included in Table~\ref{tab:relationship_categories}.

\paragraph{Social}
Peer relationships and friendships are often the most common relationship in one's social network~\citep{gorrese12peer}.
Characteristics include high levels of reciprocity~\citep{hartup99friendship}, a wide range of shared topics~\citep{hays84friendshipdevelopment} and homophily~\citep{rivas09friendship}.
Strong ties include close friends who provide emotional support~\citep{richey80bestfriends}, while weaker ties such as acquaintances or neighbors can help build connections and obtain information~\citep{granovetter83weakties}.
Several studies on online social networks have focused on the interactions of social relationships~\citep{ellison2013calling,lee09onlinecommunicationandadolescentsocialties,burke14growingcloseronfacebook}.

\begin{table}
    \centering
    \begin{tabular}{l l }
       {\textbf{Category}}  & {\textbf{Examples}} \\
        \hline
        {Social} & {best friend, neighbor, roommate} \\
        {Romance} & {dating partner, spouse, fiancé} \\
        {Family} & {parent, child, aunt} \\
        {Organizational} & {manager, colleague, pastor} \\
        {Parasocial} & {idol, fan, hero} \\
    \end{tabular}
    \caption{Examples of relationship types per category}
    \label{tab:relationship_categories}
\end{table}

\paragraph{Romance} Romantic relationships are central to adult life, leading to opportunities for intimacy and support~\citep{hartup99romanticrelationships}.
They exist in various stages such as dating, engaged and being married~\citep{stafford91romanticmaintenance, knapp80communicationrelationships}, and can develop into the formation of new families.
These relationships are often considered the closest ties.
Previous work introduced methods to classify romantic relationships in online social networks based on their network properties~\citep{backstrom14romanticpartnerships} and conversation content~\citep{tay18couplenet}.

\paragraph{Family} Family relationships are essential for building personalities and receiving social support.
Though maintained throughout lifetime, their importance may decline and are partially replaced by social and romantic ties during young adulthood~\citep{shulman75lifecycle,davidbarret16familyandfriends}.
This is reflected in contact frequency, diversity of activities and influence strength, which are lower than romantic relationships and similar to friend relationships~\citep{berscheild89rci}.
Topics shared between family relationships in online social networks typically include advice giving and household issues~\citep{burke13familiesonfacebook}.

\paragraph{Organizational} Relationships are also formed as individuals join organizations and are assigned roles within them~\citep{sluss07workrelationship,marwell70rolerelationships}.
Organizational relationships are a mixture of personal and role relationships~\citep{bridge92blendedrelationships}. This dual status leads to a stronger notion of a community or group identity~\citep{klein86workplaceidentity} and a lesser sense of trust and solidarity compared to friend relationships~\citep{myers04solidarity}.
Information exchange and politeness are expected in conversations~\citep{sias12workplacefriendships}.

\paragraph{Parasocial} The final relationship category is highly asymmetrical, consisting of celebrity-fan relationships~\citep{garimella17understandingparasocial}, involving high levels of affection from one side, resembling friendship or romantic relationships~\citep{kehrberg15iloveyoutwitter}.
Parasocial relationships are especially important to study in social networks such as Twitter, as influential figures with millions of followers can influence which topics go ``viral''~\citep{suh10rt,gayle13twitterparasocial}.

\section{Extracting Relationships}
\label{sec:datadesc}

\begin{table}

\centering

\begin{tabular}{r rrrr}

\textbf{{Category}} & \textbf{{Dyads}} & \textbf{{DM}} & \textbf{{PM}} & \textbf{{RT}}\\
\hline
{Social} & {6.6M} & {81M}  & {23.9M}  & {47M}  \\ 
{Romance} & {2.3M} & {36M} & {11M}  & {20M}  \\ 
{Family} & {324K} & {3.4M}  & {945K}  & {1.7M}  \\ 
{Organizational} & {92K} & {419K} & {316K}  & {470K}  \\ 
{Parasocial} & {360K} & {4.1M} & {3M} & {4.1M}  \\ 
\hline 
{Total} & {9.6M} & {125M} & {39M}  & {74M}  \\
\end{tabular}

\caption{Statistics of dyad pairs and tweet interactions in the final dataset for directed mentions (DM), public mentions (PM) and retweets (RT).}
\label{tab:relationship_distribution}
\end{table}

\begin{figure*}[t]
\centering
\includegraphics[width=\textwidth]{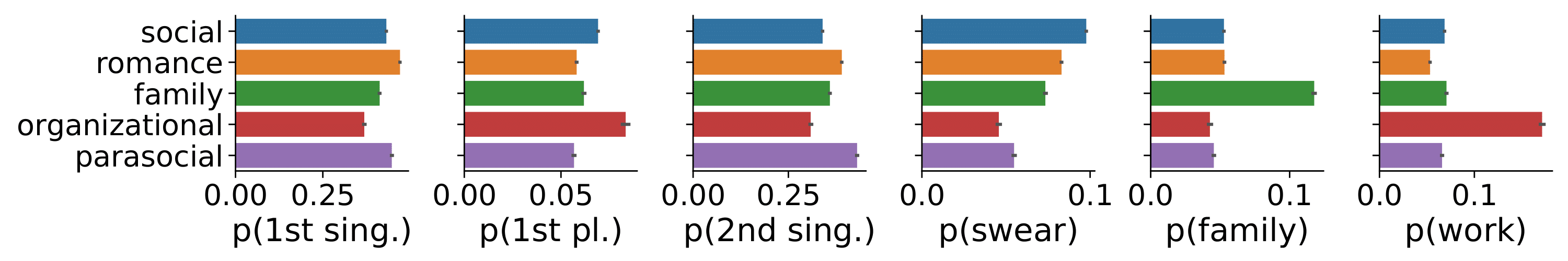}
\caption{Probability of containing a LIWC-category word in a directed mention to a specific relationship type. 
    Romance and parasocial relationships express high levels of self-disclosure by using more singular pronouns, while organizational relationships use more plural pronouns to show collective identity. Swearing is most common among social and least common within organizational relationships, possibly due to differences in social distance. Work- and family-related words are associated with the respective relationship categories. Here and throughout the paper, error bars denote bootstrapped 95\% confidence intervals.}
    \label{fig:liwc} 

\end{figure*}
In order to construct a ground truth set of dyads with labeled relationships, we use self-reported relationships between users where a user declared their relationships to another user in a tweet. 
A similar strategy has been used for extracting social roles from tweets~\citep{bergsma2013using,beller14socialroles}. We now describe the full procedure. 

We begin with a 10\% sample of all tweets posted between 2012 and 2019 and remove all non-English tweets using \texttt{pycld2}. 
We  search for all instances of the phrase `my \textbf{REL} @username', with \textbf{REL} being any string of up to three words. 
Through this search, we capture public relationship declarations such as ``My dear husband @username...".
Since this will also capture many phrases that do not correspond to a relationship, our next goal will be to filter out such instances. 

First, all phrases occurring $<$1,000 times in the dataset are removed, as we observed that most of the low frequency terms do not correspond to relationships. This process leaves 1,298 phrases that potentially map to a specific relationship.
Next, every phrase was assigned to one of the five relationship categories or labeled as invalid through a two-step annotation process by the authors. 
For each phrase, annotators were shown 50 phrases and asked to choose which categories (up to two) the phrase belonged to, if any.
To aid the annotation task, for each phrase, annotators were given five relationship-signaling tweets that used that phrase. 
Inter-annotator agreement score was measured by averaging the pairwise Fuzzy Kappa score~\citep{kirilenko16fuzzykappa}, which allows for multiple categories selected per item.
Annotators obtained a $\kappa$=0.69, indicating high agreement.
Given the high agreement, the remaining phrases were equally distributed across annotators without overlap. 

After annotation, phrases assigned either to zero or more than one relationship category were discarded (see Supplemental Material 1 for details)\footnote{Please refer below for the supplemental material}. 
Ultimately, 508 phrases were assigned to a single relationship category. 
These phrases were used to label 9,672,541 relationships between 10,410,262 users. 
Tweets were then collected for all 10.4M users from our 10\% sample from 2012--2019, totalling 238M tweets.

Three types of tweets are used in our analyses, which represent different types of interactions between users. 
(1) \textbf{Directed mentions}: Tweets where a user directs a message to a specific user by adding the username at the beginning of the tweet, typically for starting a conversation or replying to another user. 
While the mentioned user is notified that they were mentioned, this tweet does not appear on the posting user's timeline. 
(2) \textbf{Public mentions}: Tweets visible for the public audience where a username is mentioned in the middle of a tweet.
Public mentions are typically used to refer to other users, but not necessarily to have a conversation. 
(3) \textbf{ Retweets}: Instances where a user is broadcasting a tweet posted by another user. 
The number of dyads, directed mentions, public mentions, and retweets for each relationship category are shown in Table~\ref{tab:relationship_distribution}.

\section{Behavioral and Structural Differences in Relationships}
\label{sec:exploratory}

To test the quality of extracted relationships, we test communicative and network patterns in each type, validating our data using predictions from known trends for specific relationships~\citep{burke13familiesonfacebook,ellison07thebenefitsoffacebookfriends}.

\subsection{Linguistic Preferences}
Linguistic style and content reflect how an individual perceives another \citep{bell1984language}.
For instance, usage of pronouns reveal levels of self-disclosure~\citep{choudhury14selfdisclosure,wang16selfdisclosure}, and swearing terms indicate a closer social distance between the speakers~\citep{feldman17profanity}.
Comparing the use of these words by relationship category can reveal how open each relationship types are in Twitter conversations.
Using lexicons from LIWC~\citep{pennebaker2015development}, we calculate the probability of a directed mention containing one of a specific set of words: \nth{1} person singular and plural pronouns, \nth{2} person singular pronouns, and swearing terms.
Assuming that there exist topics central to a single type of relationship such as work-related topics, we also include the LIWC categories for work- and family-related words.
The results are displayed in Figure~\ref{fig:liwc}.

Communication patterns match prior expectations, as illustrated in three trends.
First, conversations in organizational relationships focus on collective identity, as shown in the highest probability of \nth{1}-personal plurals, but lowest in \nth{1}-person singular, as individuals in these relationships associate each other in the context of a larger collective entity~\citep{klein86workplaceidentity}.
Second, parasocial relationships use lesser \nth{1}-person plural pronouns but more \nth{1}- and \nth{2}-person singular pronouns, making their behavior similar to that of romantic relationships. This result is consistent with previous findings on behaviors in parasocial relationships that resemble love and affection due to the intense focus on the higher status individual~\citep{tukachinsky2010romantic}.
However, parasocial conversations also contain substantially fewer swear words compared to romantic relationships, reflecting higher perceived cost of social norm violation due to the relationships larger social distance~\citep{beers2012s}.
Also, consistent with previous findings stating the positive relationship between profanity and social distance~\citep{feldman17profanity}, swear words appear most commonly for social relationships, followed by romance and family relationships.
Finally, the figure also reveals that work- and family-related words match folk expectations: organizational and family relationships are the most likely to use their respective LIWC categories, underscoring the topical differences between relationships.
The most frequent five words for each LIWC category (shown in Table 2 in Supplemental Material) confirm that these topical-relationship interactions are not primarily driven by a single word in a category.

\subsection{Topical Diversity}
Do some relationships talk about more diverse topics than others?  
Social penetration theory predicts that the variety of topics shared in conversations should increase as relationships further develop~\citep{altman73socialpenetration}. 
We test this prediction using a topic model to analyze the diversity in communication.
Following prior work~\citep{quercia12topic}, we measure topical diversity using the entropy of the message's distribution of topics from a trained LDA model.
A 100-topic LDA model is fit using Mallet~\citep{McCallumMALLET} on a sample of 100K dyads balanced across the five categories and using five tweets per dyad to control for differences in communication frequency.
To measure diversity we calculate entropy over the mean topic distribution per dyad, then aggregate by category.

\begin{figure}[t!]
    \centering
    \includegraphics[width=0.3\textwidth]{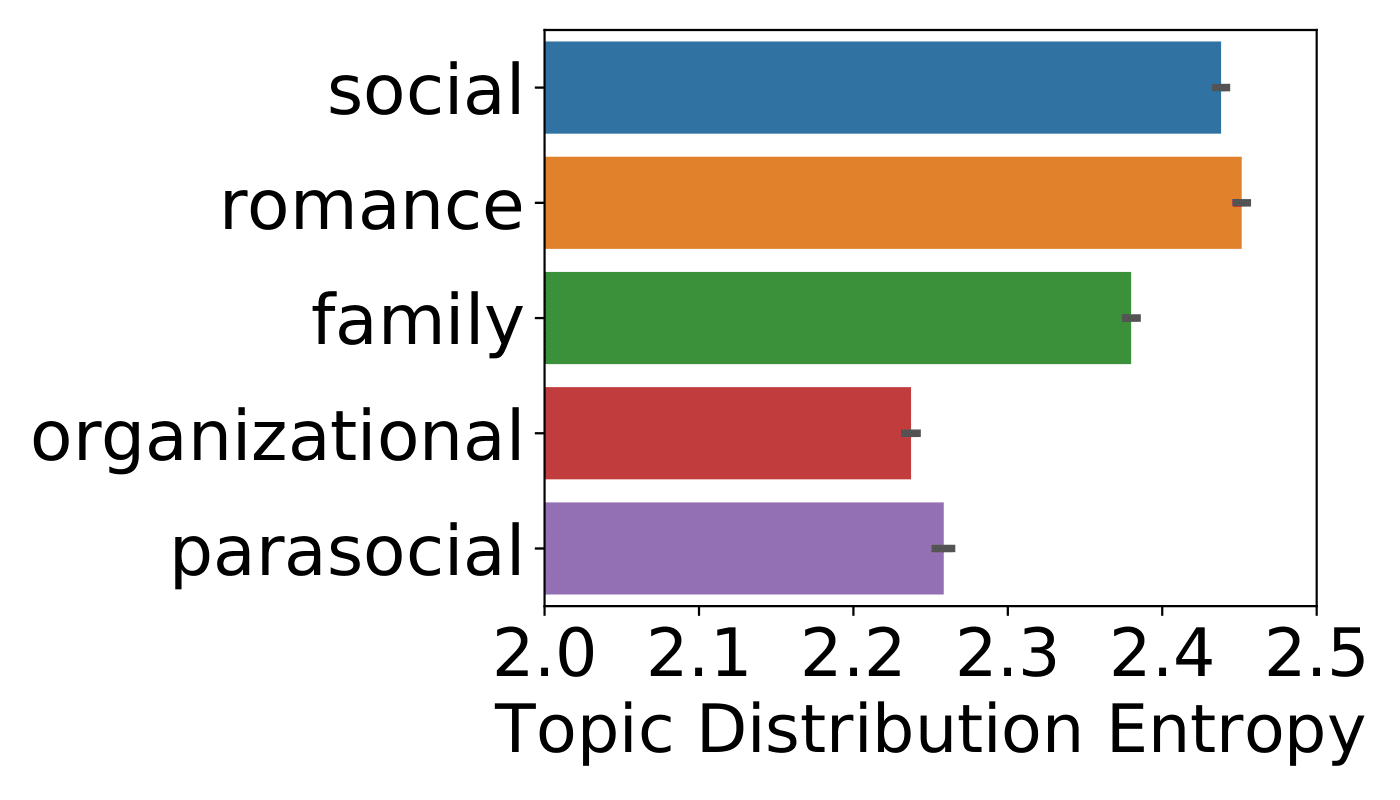}
    \caption{The average entropy of topic distributions obtained from directed mention tweets. The entropy is significantly higher for social and romance relationships, which shows these relationships contain more topics in their conversations.
    }
    \label{fig:entropy}
\end{figure}

The observed topical diversity (Figure~\ref{fig:entropy}) matches predictions from social penetration theory, with more diverse topics (higher entropy) seen in relationship categories that are more likely to contain deeper relationships with stronger ties and to have developed further such as romance, social, and family. 
In contrast, organizational relationships are less likely to communicate on topics outside their common ground~\citep{marwell70rolerelationships}.
These results were consistent over several runs with topic models trained on 20 or 50 topics.

\subsection{Network and Communication Properties}
%
\begin{figure}[t!]
    \centering
    \includegraphics[width=0.8\columnwidth]{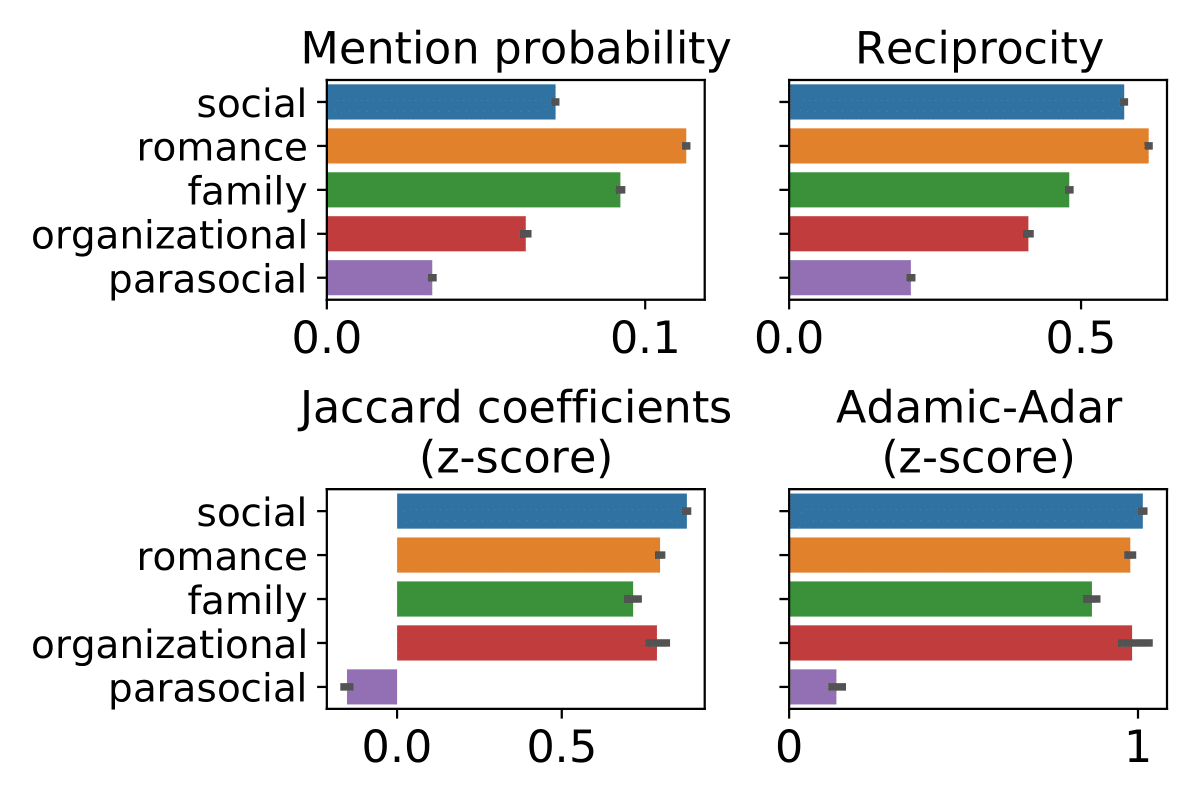}
    \caption{Network and communication features. Jaccard and Adamic-Adar scores are lowest for parasocial relationships, indicating a low similarity in neighbors of a dyad. Romance has both the highest mention probability and reciprocity, signalling the strongest level of mutual communication.}
    \label{fig:network_features}
\end{figure}
Given their different functions, relationships are expected to differ in their network and communication proprieties.
We test for these differences by examining the labeled dyads within the larger social network constructed from our entire 10\% Twitter sample. Here, two users have an edge if they both mention each other at least once, which results in a network with $\sim$1.1B edges. The dyads with labeled relationships represent a small fraction of this comprehensive social network,  as the majority of dyads have not declared a relationship. All the network properties for the dyads in our study are measured according to their users' statistics in this larger network.
Using this network and directed tweets between two people, we consider two aspects of a relationship: (1) communication frequencies and (2) the local network structure around a relationship.

\subsubsection{A comparison of communication frequencies across relationships}
Communication frequencies are measured using 
(a) the probability of a user tweeting to the other in a relationship, relative to all others in their ego network, and (b) the reciprocity in communication, measured as the ratio of tweets between two users, scaled to [0,1] where 0 indicates only one person tweets another and 1 is equal communication.
We denote \(\Gamma(u)\) as the set of neighbors of user \(u\), and \(\mathbf{{m}}_{u\to w}\) as the number of times user \(u\) mentions another user \(w\).
The probability of mentioning a specific user out of all possible neighbors is obtained as 
    \[\frac{\mathbf{m}_{u\to v}}{\sum_{w \in \Gamma(u)}\mathbf{m}_{u\to w}}.\]
We also compute the reciprocity between two users as the fraction of communications each user has made, denoted as 
    \[
        2\times \frac{\mathbf{min}\left (\mathbf{m}_{u\to v},\mathbf{m}_{v\to u}  \right )}{\mathbf{m}_{u\to v} + \mathbf{m}_{v\to u}}.
    \]
A score of 1.0 means a fully reciprocal dyad with both users communicating equally, and 0 a fully imbalanced dyad where only one mentions the other. To ensure that a relationship is valid, we only calculate reciprocity using dyads where each user has made at least one interaction with the other.

Communication frequencies, which are presented in the first row of Figure~\ref{fig:network_features}, exhibit clear differences across categories. Individuals prioritize communication within romantic relationships, consistent with prior work~\citep{Burton15romancecostly}, and have the highest reciprocity. 
High reciprocity implies two people have similar social status~\citep{verbrugge83reciprocity}; this behavior is seen most in categories likely to be between peers: romantic and social relationships.
Reciprocity levels follow expectations for differences in social status within each category, with the highest-distance parasocial relationship having lowest reciprocity.
\begin{figure*}[]
  \centering
  \subfigure[Raw frequency, category-wise]{\includegraphics[scale=0.16]{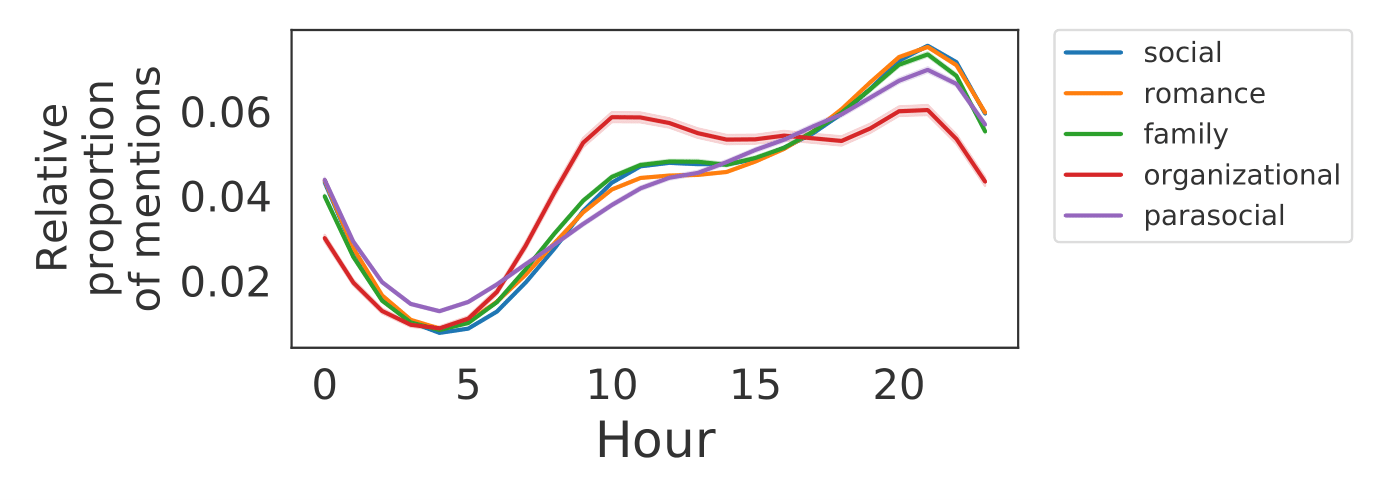}}\quad
  \subfigure[Normalized, category-wise]{\includegraphics[scale=0.16]{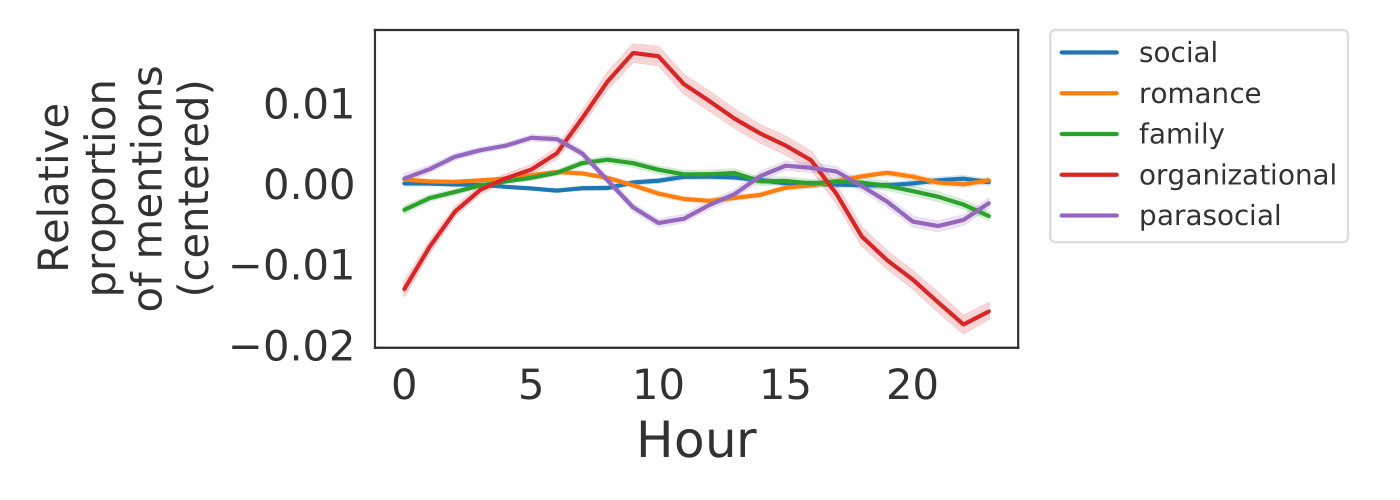}}\quad
  \subfigure[Specific to \textit{Romance}]{\includegraphics[scale=0.16]{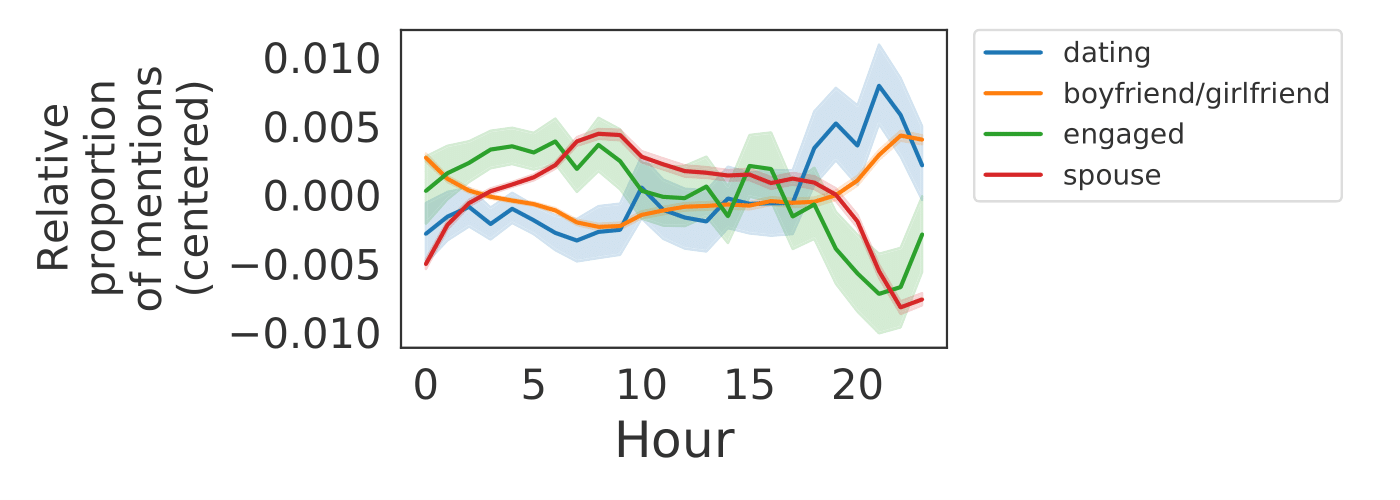}}\quad
  \subfigure[Specific to \textit{Family}]{\includegraphics[scale=0.16]{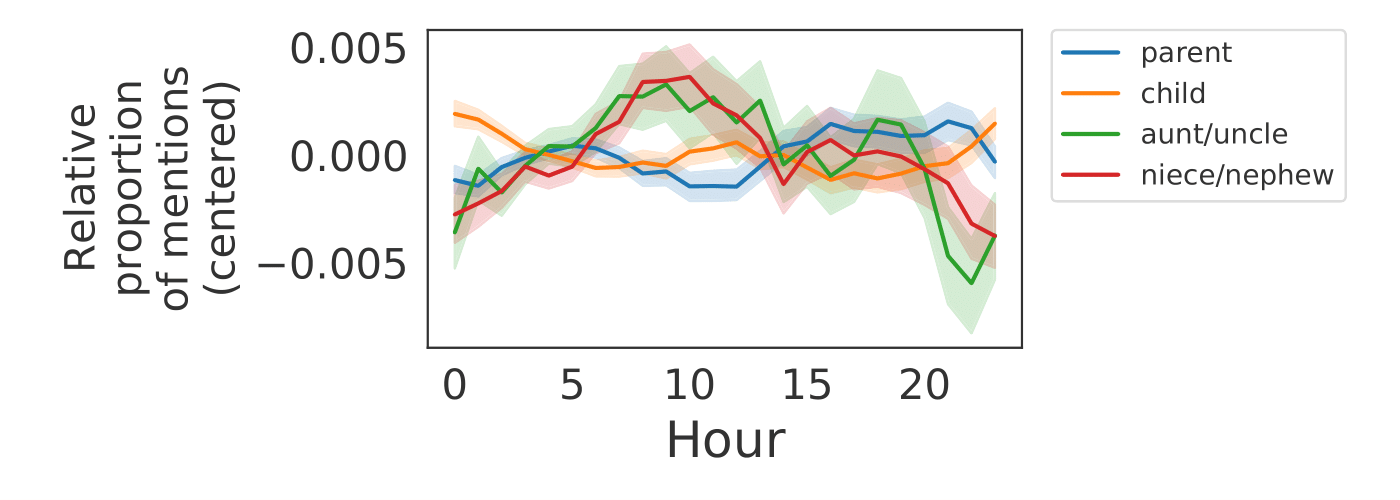}}\quad
  \caption{A comparison of mention frequency across hours of day reveal striking difference in temporal dynamics between relationship categories (a,b) and subcategories (c,d) where (b), (c) and (d) are centered relative to the mean temporal distribution across all relationship categories: (a) The un-centered communication frequency among categories (b) the centered communication frequency among categories (c) The centered communication frequency for four \textit{Romance} subcategories (d) The centered communication frequency for four \textit{Family} subcategories. Shaded regions show 95\% bootstrapped confidence intervals. }
  \label{fig:diurnal}
\end{figure*}
Reciprocity relates to status difference, where high reciprocity in contact frequency implies two users belonging to similar social statuses~\citep{verbrugge83reciprocity}.
While relationships of social and romance categories are genuinely considered as equal in status, other categories contain non-reciprocal relationships such as parent-child relationships or manager-subordinate relationships, which can explain the lower scores for family and organizational categories.
Reciprocity is lowest again in parasocial relationships, showing there exist large status differences between celebrity-fan relationships.

By observing mention probability or the likelihood to get mentioned instead of all neighboring users, we can see that romance has the highest level of relative importance by far, while, surprisingly, social relationships drop to the level of other types.
This result is consistent with findings showing that individuals prioritize communicating with their romantic partners over other relationships~\citep{roberts11communicationinsocialnetworks}.

\subsubsection{A comparison of network properties across relationships}
We also consider two types of network properties: (a) the Jaccard Index of the two users' friends and (b) Adamic-Adar index~\citep{adamic03adamicadar}, both frequently used for measuring the likelihood of an edge between two nodes. To allow for direct comparisons among dyads, we use the z-normalized score for each metric instead of the raw score as  %
    $    zscore(u)=\frac{x-\mu}{\sigma},$
where \textit{x} is the raw score, \(\mu\) and \(\sigma\) are the mean and standard deviation computed from the neighboring dyads of \textit{u} other than \textit{v}. Supplemental Material \S3 provides a longer explanation of how these values are computed.

The patterns in network structure (Figure~\ref{fig:network_features} bottom) also match expectations. First, parasocial relationships exhibit very low Jaccard index and Adar-Adamic scores. This is expected as celebrities are embedded in very different social structures from that of their fans and do not have many connections in common. Second, the \textit{family} category has significantly lower Jaccard index and Adar-Adamic score than the social, romance, and organizational categories. This is likely due to two reasons: First, unlike social, romance, and organizational relationships, family relationships do not depend on network structure. Indeed, a family relationship is established regardless of social proximity. However, social, romance, and organizational highly depend on social proximity as these relationships are established through social mechanisms such as friends introducing their friends to each other (i.e. triadic closure) \citep{kossinets06empirical}. Second, family ties tend to be well embedded within family networks (e.g. siblings may have other common family connections), but also tend to be much smaller than other relationships such as social and organizational. Due to these differences in volume, family ties are overall less embedded.

\subsection{Diurnal Communication Patterns}

Individuals manage their communications differently according to social relationships or ``identities'' such as \textit{social}, \textit{work} and \textit{family}~\citep{ozenc11lifemodes,min13miningsmartphonedata}. By observing diurnal Twitter usage patterns through the timestamps of tweet messages~\citep{golder2011diurnal}, we show the existence of both between- and within-category communication differences across relationship types. 

Using our massive volume of communications, we compute a distribution representing the fraction of messages exchanged for each relationship dyad during each hour of a day. With the number of mentions from user \textit{u} to user \textit{v} as \(\mathbf{{m}}_{u\to w}\), we bin each mentioning tweet according to the hour of the day it was created. We then define \(t_{u\to w}\left ( i \right )\) as the fraction of mentions produced in the \textit{i}-th hour so that \(\sum_{i=0}^{23}t_{u\to w}\left (i \right )=1\). 
We restrict our analysis to tweets where a local timezone of the tweeting user is provided along with its global timestamp and convert the tweet to its local time. Also, we only consider cases where the sending user has made at least 5 activities for better smoothing of the distribution.

After we compute the diurnal communication distributions for each dyad, we can aggregate across different relationship categories or subcategories to obtain category-wise diurnal distributions. 
We provide a comparison of the diurnal distributions aggregated across different categories, shown in Figure~\ref{fig:diurnal}(a). While all categories share the same pattern of substantially lower communication during dusk and peaks around evening, similar to previous work~\cite{golder2011diurnal}, we can observe slight differences around daytime.
To further examine such differences, we center the distributions by subtracting each with a global mean \[t_{global}\left ( i \right )=\frac{1}{\left | S \right |}\sum_{j=1}^{\left | S \right |}t_{\left ( u\rightarrow v \right )_j }\left ( i \right ),\left \{ \left ( u \rightarrow v \right )_j \in S\right \}\]
with \(S\) as the set of all dyads across all categories, or in Figures \ref{fig:diurnal}(c) and (d), the different subcategories we consider.
The centered result (Figure~\ref{fig:diurnal}(b)) shows a notably higher communication rate for the \textit{Organization} category during work hours (9-16) which drops afterwards, possibly due to moving away from a work activities and towards friends and family chatter~\citep{farnham11facetedidentity}.

Due to our data scale, we can examine communication patterns \textit{within} the same relationship category. The trends for the \textit{Romance} category, shown in Figure~\ref{fig:diurnal}(c), reveal that even within romance relationships,  diurnal patterns can be partitioned into early (``dating'' and ``boyfriend/girlfriend'') versus stable (``engaged'' and ``spouse'') stages, where the former communicates more during late hours. One possible reason is that the latter group may consist of more married couples that share the same physical space during the evening and have fewer reasons to communicate through Twitter. Within family relationships (Figure~\ref{fig:diurnal}(d)), we show that aunt/uncle - niece/nephew communication is 
more intense during the day compared to parent-child communication. We conjecture that this tendency reflects the lesser degree of perceived closeness between extended families as opposed to direct kin, restricting communication during late hours.

\section{Relationship Classification}
\label{sec:classification}

Given that relationship types differ in linguistic preferences, topic diversity, and network properties, we now test whether they can be accurately classified from these features.

\subsection{Task and Experimental Setup}
\label{sec:classifier-setup}

We classify a dyad into one of the five relationship categories on the basis of its behavioral and communication features.
The prediction task is conducted with both balanced and imbalanced datasets.
{\bf Balanced set:} Training data uses 200K dyads per category, randomly upsampling  organizational relationships which had fewer samples. A random sample of 2K dyads were used for validation data. 
The test set contained 17,522 dyads per category, where all classes were downsampled to match the least common class, organizational relationships.
{\bf Imbalanced set:} 2M dyads are randomly selected and split by a 8:1:1 ratio into training, testing and validation partitions. In both settings, users were constrained to be in only one partition. 
We ensure that at least one user of the dyad performs at least five interactions, in order to sufficiently represent a dyad's interactions. To control for differences in communication frequency, we restrict the data to at most 15 tweets from each user in a dyad, keeping up to 5 tweets for each type of communication: directed tweets, retweets, and public mentions. To avoid data leakage, all tweets that contained the phrase used for labeling the relationship type were removed prior to this process.

\subsection{Proposed Models}
\label{sec:model}

To capture information from the different types of communication, we introduce a new deep learning model that performs multi-level encoding to represent these texts. Our base architecture builds upon the model of  \citet{huang-carley-2019-hierarchical} using the parameters of the RoBERTa pre-trained language model \citep{roberta}. Each tweet and bio text for a dyad are encoded as constant-length vectors by encoding each with RoBERTa and mean pooling the output layer's word piece embeddings for the content.
These pooled content encodings are fed into a separate 6-layer Transformer network \citep{vaswani2017attention}; this second level of encoding summarizes the different sources of text information through its attention mechanism. The output layer of this second-level model is mean pooled as a representation of 
all communication.

The final dyad representation concatenates the multi-level encoding with (i) character-level embeddings for the username created using 1-dimensional convolutional filters~\citep{kim1dcnn} and (ii) the four network statistics described in Section \ref{sec:exploratory}. This representation is fed through two linear layers using ReLU activation with dropout before a softmax is applied to classify the dyad. RoBERTa models were first pre-trained using 3M training set tweets; then the full classification model was trained end-to-end. Supplemental Material 4.1 provides details of all hyperparameters and training procedures.

\noindent\textbf{Baselines}
We introduce two baselines: the first is a random guess for all samples, and the stronger second baseline combines text and network features and uses them to train an XGBoost~\citep{xgboost} classifier.
Uni-, bi-, and tri-grams with more than 10K frequency are used as features.
We also add features for frequencies in category of the LIWC and Empath~\citep{empath} lexicons.
Network features are identical to the neural model. Details are included in Supplemental Material \S4.2.

\begin{table}[t]
    \centering
    \resizebox{0.48\textwidth}{!}{
    \begin{tabular}{ l r ccccc c}
        & {\textbf{Model}}
        & {Soc.} & {Rom.} & {Fam.} & {Org.} & {Para.} & { F1 }\\
         \hline 
          \parbox[t]{2mm}{\multirow{3}{*}{\rotatebox[origin=c]{90}{Bal.}}}
        & {\textit{Random}} & {0.20} & {0.20} & {0.20} & {0.20} & {0.20} & {0.20} \\
        & {GBT Model} & {0.45}& {0.57} & {0.55} & {0.64} & {0.55} & {0.55}  \\
        & {Our Model} & {\textbf{0.60}} & {\textbf{0.69}} & {\textbf{0.69}} & {\textbf{0.79}} & {\textbf{0.72}} & {\textbf{0.70}}  \\
        
        \hline
        \parbox[t]{2mm}{\multirow{3}{*}{\rotatebox[origin=c]{90}{Imbal.}}}
        & \emph{Random} & {0.62} & {0.30} & {0.05} & {0.02} & {0.10} & {0.20}  \\
        & \emph{Maj. Class} & {0.76} & {0.00} & {0.00} & {0.00} & {0.00} & {0.15}  \\
        & {GBT Model} & {0.80}& {0.52} & {0.33} & {0.28} & {0.28} & {0.44}  \\
        & Our model & {\textbf{0.84}} & \textbf{0.68} & \textbf{0.51} & \textbf{0.50} & \textbf{0.38} & \textbf{0.58} \\
    \end{tabular}
    }
    \caption{Performance comparison for different settings. The F1 score is used to measure the performance in all cases. The first five columns show the F1 scores measured only from samples whose ground truth label belongs to each category as a binary classification task. The last column is the combined Macro F1 score computed as a multi-task classification task.}
    \label{tab:relationship-prediction-performance}
\end{table}

\subsection{Results and Findings}
Our model can accurately recognize different relationships, attaining a macro F1 of 0.70 on the balanced dataset (Table \ref{tab:relationship-prediction-performance}), indicating that relationship categories are identifiable from their network and communication patterns. This performance substantially improves upon that of the XGBoost baseline and random chance.
In the balanced setting, the model is most accurate at predicting organizational and parasocial and least at social relationships. We attribute this difference to the intra-class diversity; while organizational and parasocial relationships typically have narrowly-exhibited behavior (e.g., low topic diversity for organizational in Figure~\ref{fig:entropy}), social relationships can take many forms, e.g., friends, neighbors. This diversity likely makes the class harder to distinguish.

In the imbalanced setting that reflects the natural distribution of classes, our model offers an even larger performance improvement over baselines. Here, most dyads have a social relationship (76\%; cf. Table~\ref{tab:relationship-prediction-performance}), yet the model is still able to reliably identify all classes.
This result highlights the applicability of the model in
real-world settings, which is crucial for studying communication dynamics.

Separate ablation studies were performed to test the information from each type of communication (direct, public mention, or retweet) and for the addition of user and network features. 
Among communication types, public mentions provided the most information (highest performance) for predicting relationship types (0.56 in balanced setting). We speculate that a user has to include context about the mentioned user and their relationship when mentioning them in a public tweet, to provide an explanation to the audience. This information is not required in directed mentions since the expected audience is only the two users.
The addition of user features and network features to text features also significantly improves performance, most notably for parasocial relationships, where adding user profiles and network features boosts the F1 score by 0.13.
The significantly lower Jaccard coefficients and Adamic-Adar scores of dyads in the parasocial category~(Figure~\ref{fig:network_features}) likely makes it easier to identify the relationship type, when incorporated as features.
Table with full results is in Supplemental Material (Table 3).

\subsection{Testing the Validity of Classifier Models}
As a further validation of the relationship classification model, we test whether the relationships inferred by our model mirror the behavioral properties seen in the labeled relationships. A random sample of 1M dyads is collected where one user has made at least five interactions, mirroring our classifier setup (\S\ref{sec:classifier-setup}). 
We then collect mention, reply, and retweet activities made between both users of the dyad, and apply the classifiers on these new users.
As temporal information is not used in the classifier but shows clear differences by relationship (Figure~\ref{fig:diurnal}), we test whether the communication patterns in these inferred relationships are similar. 
The resulting time series from inferred users were highly correlated with those of the labeled data correlations, ranging from 0.928 to 0.947, shown in detail in Supplemental Table 5 and visualized in Supplemental Figure 1. This result indicates the model infers relationships that have highly similar behavior to the labeled data in practice (despite not being trained on these features), e.g., dyads in the organizational category focusing more of their communication during daytime and dropping in volume after work hours  for both labeled and inferred data.

As a second test of validity, we examine the distribution of inferred relationships in the random sample. The resulting distribution differs substantially from that of the labeled data, as expected: Social 23\%, Romance 23\%, Family 14\%, Organizational 5\%, and Parasocial  36\% (cf.~Table~\ref{tab:relationship_distribution}).
In the random sample we observe more Parasocial, which aligns with earlier expectations that Twitter is largely a mass media platform rather than a social network \citep{kwak2010twitter}. However, unlike expectations from earlier work, we observe a significant uptick in stronger social relationships: Social and Romance together account for $\sim$46\% of the random sample. We attribute this result, in part, to the requirement that dyads in the random sample must have at least five directed tweets, which likely increases the presence of stronger ties who are more likely to talk more. Our results point to the social nature of Twitter and the need for future work to examine how all relationships are manifested on Twitter---not just those that communicate---to establish to what degree the platform now serves as a social network.

\section{Retweet Prediction with Relationships}
\label{sec:prediction}

Retweeting is central to information spread on Twitter. Given its significance, several studies have introduced approaches to model and predict this behavior.
Factors identified include: content-based features (e.g., hashtags, URLs~\citep{bakshy2011influencer}), network features (e.g., tie strength~\citep{yuan16rtreply}), and user popularity~\citep{hong11popular}.
Despite all these efforts, retweet prediction remains a notoriously hard problem~\citep{martin16limitsretweet}. Our findings in Section \ref{sec:classification} suggest one potential way to improve past efforts.
The fact that relationships can be predicted based on tweet conversations indicates a promising connection between the topic of a tweet and the type of relationship that would have interest in it. 
This leads us to hypothesize that a user's probability of retweeting depends on the interaction between the tweet's content and the relationship to the tweet's author.
As such, we test whether incorporating the relationship type between two users $u$ and $v$ can improve the prediction accuracy for whether user $v$ will retweet a particular tweet by $u$.

\subsection{Dataset}
We first select the same number of dyads per relationship category to balance across different relationship types. For each dyad we collect all retweets that occurred between the two users, but remove instances where user mentions occurred in the original tweet. While being mentioned in a tweet is a strong motivator for a retweet to occur~\citep{jenders2013analyzing}, here our goal is to understand the interplay between the content of the tweet and the relationship properties of the dyad and thus remove tweets with mentions.

We focus on a balanced prediction task where for each positive retweet that occurred between a dyad, we assign one tweet produced by the same user around the same time which did not get retweeted by the other user. As a result, 50,000 positive and negative tweets were sampled per category and split into training, validation, and test sets at a ratio of 8:1:1.

\begin{table}
    \centering
    \begin{tabular}{lrrrrr}
    {Model}  & {Soc.} & {Rom.} & {Fam.} & {Org.} & {Para.} \\
    \hline
    {Baseline} & {0.61} & {0.63} & {0.60} & {0.64} & {0.61}\\
    {With relationships} & \footnotesize{0.64} & {0.65} & {0.63} & {0.64} & {0.62} \\
    \end{tabular}
    \caption{Classwise F1 performance comparison of retweet prediction task on a balanced dataset. The presented order is Social, Romance, Family, Organizational, and Parasocial.}
    \label{tab:retweet_categories}
\end{table}
\subsection{Models for Retweet Prediction}
The models used for the prediction task are also based on a pretrained RoBERTa model for text classification.

\paragraph{Baseline model} encodes the tweet text using RoBERTa and preserves the embedding of the first position corresponding to the [CLS] token, which is a common practice in BERT-based classification tasks~\citep{devlin2018bert}. Features for the baseline model are the contents of the tweet and the number of followers  (log-scaled) and the existence of a URL, both frequently used features in retweet prediction~\citep{petrovic11rt,suh10rt}. The [CLS] embedding is passed through a linear layer and is concatenated with sparse features: the log-scaled number of followers and existence of a URL. This concatenated embedding is passed through two additional layers to be transformed into a single scalar value, where we apply a sigmoid function to convert into a score between 0 and 1, with 1 indicating a potential retweet.

\paragraph{Relationship-aware Model} extends our baseline model using a number of representations for relationship information. We first use a direct encoding of the relationship category into a 256-dimensional vector which is trained along with other parameters. The textual information of the declared phrase associated with the relationship type (e.g., ``my \textit{best friend}'') is also encoded using a character-level CNN with 1-d convolution~\citep{kim1dcnn} then max-pooled, resulting in another representation vector of size 768. This number is obtained by using 256 convolutional filters each for kernels with sizes of 3,4, and 5. Finally, we add the relationship category as one-hot features in addition to the other sparse features, and concatenate them with the dense embeddings.

We set $d$ to 768 and the learning rate to 1e-6.
All models are trained with batch sizes of 16 and for a maximum of 10 epochs. We select the model with the highest validation F1 score, which is computed for every 5000 steps.

\subsection{Results}

\begin{table}[t]
    \centering
    \resizebox{0.48\textwidth}{!}{
    \begin{tabular}{ l cccccc}
        &  \multicolumn{3}{c}{{{without URLs}}} &  \multicolumn{3}{c}{{{with URLs}}}\\
        {\textbf{Model}}  & {Pre.} & {Rec.} & {F1} & {Pre.} & {Rec.} & {F1} \\
         \hline 
        {Baseline} & {0.58} & {0.69} & {0.63}  & {0.53} & {0.85} & {0.65} \\
        {With relationships} & {0.58} & {0.71} & {0.64} & {0.53} & {0.87} & {0.66} \\
    \end{tabular}
    }
    \caption{Model performance (Precision, Recall, F1) at predicting retweets of messages with or without URLs. The addition of relationship types leads to an increase in the F1 score by boosting recall.}
    \label{tab:retweet-prediction-overall}
\end{table}

Adding relationship information improves performance for retweet prediction, as shown in Table~\ref{tab:retweet-prediction-overall}, which is known to be a difficult task~\citep{martin16limitsretweet}. Incorporating relationship types provides a 1\% performance increase in F1 score due to an increase in recall by 2\%.  Here, we show separate performances for test data tweets with and without URLs; tweets with URLs are likely to have different retweet dynamics on the basis of the content in the URL (e.g., retweeting a linked news story versus a personal message) and the different social uses of retweets \citep{boyd10retweet}.

\begin{figure}[t!]
    \centering
    \includegraphics[width=1\columnwidth]{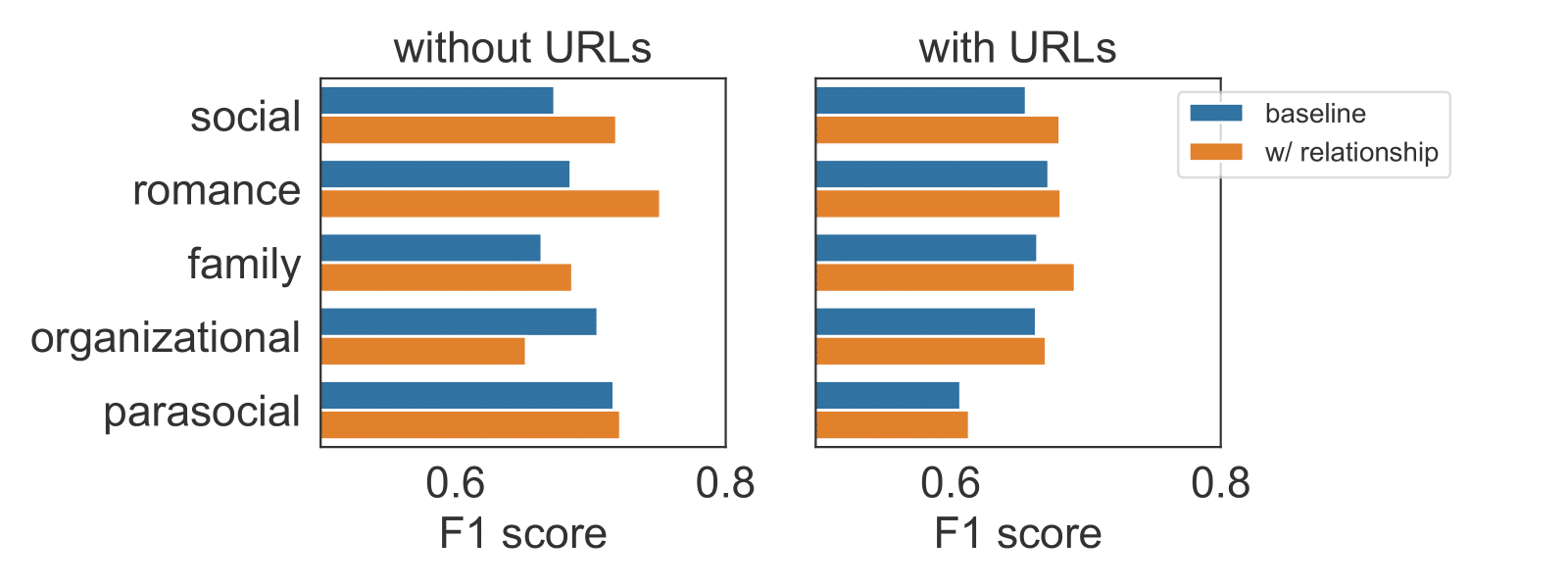}
    \caption{A comparison of baseline vs. relationship-infused models for retweet prediction on tweets with and without URLs. In both settings, the addition of a relationship type improves the predictive performance.}
    \label{fig:rt_comparison}
\end{figure}

Analyzing performance improvement by relationship type reveals a more complex picture of improvement, shown in  Figure~\ref{fig:rt_comparison}. 
For tweets containing URLs, the addition of relationship information consistently improves performance, while we observe the largest performance gains for tweets without URLs. In particular, in tweets without URLs, the model sees increases of 2.2\%, 3.1\% and 2.3\%  for social, romance, and family categories  respectively, signalling that the model can use this social information to decide where a person in that relationship is likely to retweet based on the content.
However, the model performs worse for predicting retweets from organizational relations (2.8\% decrease) which lowers the overall performance reported in Table \ref{tab:retweet-prediction-overall}. In the case of tweets containing URLs, F1 scores increase across all categories, with social (3.8\%) and family (4.2\%) categories benefiting from the largest gains.

We observe that the larger gains seen in social, romance, and family relationships are due to increased recall (see Table 4 in Supplemental Material). This increase suggests that individuals embedded in social, romance, and family relationships retweet content that is less likely to be retweeted normally (e.g., mundane personal events) because of the nature of the relationship, which the relationship-aware model is able to use to correctly identify the content will be retweeted. This result is further evidence of the interaction between communication patterns and relationship types.

\section{Ethics and Limitations}

\paragraph{Ethical Considerations}
This research was performed on only public data, in accordance with the Twitter Terms of Service. However, users  do not necessarily expect their data to be collected for research purposes nor to be disclosed~\citep{fiesler18twitterresearchethics}. Furthermore, the potentially-sensitive nature of interpersonal relationships being revealed necessitates that additional privacy steps must be taken and that benefits must outweigh harms.
To mitigate risk to individuals, we report only aggregate information and focus on broad effects, avoiding any focus on marginalized groups or sensitive relationships \citep{townsend2016social}. Further, data and models will only be shared upon confirmation of ethical principles of use.
Counterbalancing these risks, this study offers substantial benefit to our understanding of social processes and how relationships influence what we share and what we hear about. As demonstrated in Section \ref{sec:prediction}, understanding these relationships can improve the algorithms that individuals come into contact with regularly, such as social and content recommendation systems.
\paragraph{Limitations}
Data used in this study relies on self-reported, public declarations, which may not occur across all dyads in Twitter. The willingness to declare such relationships is likely indicative of a stronger tie between users. Therefore, our work may not reflect the behaviors of people in relationships less likely to be declared. Such non-declarations can be due to a variety of reasons such as increased desire for privacy, weaker association of that relationship type, or even the potential social stigma around declaring the relationship. Further, our data and model depend on observing communication between two users; as not all users with a meaningful social relationship also communicate on Twitter (e.g., spouses on Twitter who talk offline), our model is unable to identify such relationship. While the results of our work are largely in line with prior expectations from sociology and psychology, future work is needed to understand what biases, if any, stem from using only self-reported relationships and how these relationships fit within the broader space of relationships that exist between users on Twitter. Finally, our validation efforts are at the population level and future work could  perform additional validation on specific relationships through crowdsourced labeling of both the self-declared and the inferred relationships obtained through our model.

Our analysis is based on a 10\% sample of Twitter data and while consisting of billions of tweets, this sampling by nature omits tweets between users that would affect the inference of network and communication statistics. Specifically, while the measured using the best-available data and follow past scholarship that used similar datasets to infer network properties~\citep[e.g.,][]{bliss2014evolutionary,pierri2020topology}, our results likely underestimate the presence of edges and rates of communication frequency due to sampling.

The current work makes two simplifying assumptions about a dyad's relationships: (i) the relationship is of only one type and (ii) does not change over time. In practice, relationships evolve over time and categories such as social and organizational can indeed overlap. Our simplifying assumptions allow us to perform these initial studies at large scale. However, future work could relax these constraints with sufficient longitudinal data or with additional self-reported data. 

\section{Related Work}
Due to the difficulty of collecting ground truth data, only a handful of studies have examined real social relationship of different categories on social media. \citet{min13miningsmartphonedata} conduct a survey on 40 participants to obtain their SMS data and categorize their contacts into \textit{family}, \textit{work}, and \textit{social}. The authors show that it is possible to infer these relationship types using features such as the geographical similarity between two users and their contact patterns. More recently, \citet{Welch2019LookWT} collected private data for the ego network of one user and 104 alters where there exact relationship was known; the combination of text and behavioral data (such as message duration) were used to predict relationship properties.
Other studies have predicted the existence of a specific relationship from interpersonal interactions, the most common being a romantic relationship. \citet{backstrom14romanticpartnerships} show that dispersion---a metric which the authors introduce for measuring how well-connected mutual neighbors are---is a better predictor of romantic relationships than embeddedness within the Facebook network. Similarly, \citet{tay18couplenet} introduce a model that compares the similarity of messages shared between two people on Twitter to predict whether they are in a romantic relationship. 
Our study goes substantially beyond these studies by simultaneous testing a comprehensive set of relationship types and examining orders of magnitude more data for each relationship type.

Rather than study categories of relationships, some studies instead measure attributes of the relationship themselves, e.g., their relative status. \citet{adali12relationships} use social and behavioral features derived from Twitter activities to distinguish between relationship groups defined by their word usage and a clustering algorithm. \citet{gilbert09tiestrength} and \citet{gilbert12tiestrength} use interaction-based features provided by the Facebook platform to predict the tie strength of user dyads in Facebook and Twitter. \citet{rashid2017dimensions} and \citet{rashid18rel} use a conversation dataset from the TV series \textit{Friends} to label the relationships between the main characters with properties such as ``equal-hierarchical'' or ``pleasure-task oriented'', then show that these properties can be predicted using text features. \citet{choi20socialdimensions} predict the prevalence of specific social dimensions such as trust or power differences using deep learning classifiers on Twitter interactions between two users.
These studies offer a complementary view of relationships and our introduction of a new large-scale dataset with relationship categories opens up new future work for testing how these attributes align or differ across  categories.

Social networking platforms serve multiple roles as both social and informational networks \cite{arnaboldi13dynamicsofpersonalsocialrelationships}, and as such individuals may form ties with others for different purposes.
Several works have examined how different properties of the dyads reflect their information sharing behavior. In particular, studies have focused on the characteristics of interactions between close users, showing that such dyads (i) have a greater tendency to share less common hashtags which may indicate community belonging~\citep{romero13topical}, (ii) are more likely to share content than with strangers~\citep{quercia12topic,bakshy2009social}, and (iii) frequently engage through actions such as mentioning each other or sharing posts~\citep{jones13inferringtiestrength}. Other work focuses on user interactions with influential users such as celebrities or politicians who possess a large follower base. These users gain widespread influence by specializing on narrow topics~\citep{cha2010measuring} or posting messages with strong sentiments~\citep{dang2013investigation}.  While our study also aims to identify communication and interaction differences between different types of user relationships, we go beyond the widely studied themes, such as close ties or influential users, to study how specific types of relationships interact with information sharing, showing differences in which content is communicated across specific types (e.g., Figure \ref{fig:liwc}) and that knowing the relationship type aid in predicting which content is retweeted.
\section{Conclusion}
Not all ties are equal: friends, family, and lovers all have different social, linguistic, and temporal behaviors---yet, social network studies have typically limited themselves to networks with edges that encode only the existence of a relationship, but ignore the \emph{type} of that relationship. Using a dataset consisting of the interactions between 9.6M dyads on Twitter with known relationship types, we introduce a new approach that explicitly models interpersonal relationship types in social networks. We make the following three contributions towards understanding relationships in networks. First, we show that the linguistic, topical, network, and diurnal properties in online communication between different relationship types match predictions from theory and observational studies. Second, we demonstrate that relationship types can be accurately predicted using text and network features combined with state-of-the-art deep learning models. Third, we show that knowing relationship types improves performance when predicting retweets---demonstrating the benefits of predicting relationships at scale. The addition of relationship type significantly improves the recall of retweets for social, family, and romance relationships, which are considered more personal.

Our proposed approach, combined with the consistency of our results with existing literature on social relationships, further demonstrates the value in studying social media networks to further understand the differences in communicative behaviour across interpersonal relationships.
Furthermore, as evident from the performance of our relationship classifier and the improvement on retweet prediction, our work enables new types of analyses that benefit from large-scale relationship-aware networks such as modeling network evolution, information diffusion dynamics, and community structure. Overall, our work provides a stepping stone towards incorporating relationship types in several research questions in social and network sciences.

\section*{Acknowledgments}
The authors thank the reviewers for their thoughtful comments and Carol Zheng for her feedback. 
This material is based upon work supported by the National Science Foundation under Grants No 1850221, 2007251, and 1815875, by the Volkswagen Foundation, and by the Air Force Office of Scientific Research under award number FA9550-19-1-0029.

\bibliography{main.bib}

\clearpage
\appendix
\renewcommand\thefigure{\thesection \arabic{figure}} 
\renewcommand\thetable{\thesection \arabic{table}}   
\setcounter{figure}{0}
\setcounter{table}{0}
 
\begin{appendices}

\section*{1. Further details on filtering user dyads}
\label{sec:detail_filtering}
To maintain precision, phrases that could signal multiple relationships, e.g., \textit{bro} for a biological brother or a friend, were removed from our dataset.
The removal of such relationships is an attempt to preserve the distinctiveness of relationships at the expense of sample size.
Additionally, for dyads where both users declared different relationships or different relationships were declared several times in the same dyad, we randomly sampled one instance and dropped the remaining duplicates.
Another filtering step was to remove personal relationships declared towards public figures.
Parasocial celebrity-fan relationships often entail a degree of affection from one side that may look like friendships or romantic relationships~\citep{dibble16parasocialinteraction,kehrberg15iloveyoutwitter}.
For instance, the account for Justin Bieber was declared as a boyfriend by 3,749 distinct users.
We removed all declarations of non-parasocial relationships that was targeted to Twitter accounts with more than 10,000 followers, which is a reasonable threshold for identifying influential users.

\section*{2. Further details on LIWC analysis}
\label{sec:detail_liwc}
Although we have shown that each relationship category has distinct linguistic properties in conversations through tweets which are visible through different levels of LIWC category words used, it is also worth knowing which words propel such differences across relationship categories. For this reason, we provide the top-5 words for each LIWC category that appear most in each relationship category and list their percentage over the LIWC category words, which is shown in Table~\ref{tab:frequent-liwc-words}. For instance, the top-5 swear words used in the social category account for roughly 53\% of the total count of swear words. THe percentage values do not indicate volume but how evenly distributed the words in a LIWC category are. We can observe that for swear words, the distributions are relatively the same across different relationship categories. By combining this with the results shown in Figure 1 in Section 5.1., we know that while there is only a small difference in which words to use for swearing, having relationships belonging to certain categories such as social greatly increases the probability of including it in a message.

\begin{table}[t]
    \centering
    \begin{tabular}{p{2.4cm}p{4.4cm}}
         \small{\textbf{Feature}} & \small{\textbf{Description}} \\
         \hline
         \small{\textit{Conversation}} \\
         \small{Directed mentions} & \small{Tweets and replies that are directed to a single user}\\
         \small{Public mentions} & \small{Tweets broadcasted to one's follower network that mentions a specific user}\\
         \small{Retweets} & \small{Messages retweeted from a specific user}\\
         \\
         \small{\textit{User information}} \\
         \small{Description} & \small{The description text in a user's bio} \\
         \small{Username} & \small{The username associated to a user's account} \\
         \small{Display name} & \small{The name displayed in front of the username in a tweet} \\
                        \\
         \small{\textit{Network}} \\
         \small{Adamic-Adar} & \small{The Adamic-Adar score between two users}\\
         \small{Jaccard} & \small{The Jaccard coefficient between two users}\\
         \small{Mention proportion} & \small{The relative mention importance score, applied for both directions}\\
    \end{tabular}
    \caption{Types of information used for relationship prediction task}
    \label{tab:features}
\end{table}

\begin{table*}[t]
    \small
    \centering
    \begin{tabular}{lcccccc}
        & \multicolumn{6}{c}{{\textbf{Word (Proportion in LIWC category)}}}\\
         {\textbf{LIWC category}} & {Order}
        & {Social} & {Romance} & {Family} & {Organizational} & {Parasocial} \\
         \hline 
         \textbf{{LIWC: swear words}} \\
          & {1} & {shit (17.31)} & {shit (14.43)} & {shit (16.84)} & {shit (17.05)} & {shit (12.68)} \\
          & {2} & {fuck (10.94)} & {ass (12.73)} & {ass (11.07)} & {fuck (11.3)} & {fucking (9.85)} \\
          & {3} & {ass (10.36)} & {fuck (10.37)} & {fuck (11.05)} & {ass (10.19)} & {fuck (9.21)} \\
          & {4} & {bitch (9.13)} & {bitch (9.11)} & {bitch (8.37)} & {bitch (6.87)} & {ass (8.87)} \\
          & {5} & {damn (5.86)} & {damn (6.04)} & {damn (5.73)} & {damn (6.35)} & {damn (8.26)} \\
         \textbf{{LIWC: family-related}} \\
          & {1} & {bro (22.69)} & {baby (41.37)} & {mom (9.89)} & {bro (22.18)} & {baby (32.67)} \\
          & {2} & {baby (12.25)} & {mom (6.68)} & {baby (9.27)} & {baby (10.92)} & {bro (9.28)} \\
          & {3} & {mom (8.12)} & {dad (3.14)} & {son (8.43)} & {family (5.04)} & {family (8.55)} \\
          & {4} & {fam (4.28)} & {bro (3.11)} & {dad (8.1)} & {brother (4.9)} & {mom (6.19)} \\
          & {5} & {dad (4.24)} & {daddy (2.84)} & {bro (7.97)} & {fam (4.77)} & {brother (2.72)} \\
         \textbf{{LIWC: work-related}} \\
          & {1} & {work (11.91)} & {work (12.97)} & {work (10.72)} & {work (8.65)} & {read (10.44)} \\
          & {2} & {school (7.47)} & {school (8.03)} & {school (6.97))} & {team (3.31)} & {work (6.27)} \\
          & {3} & {read (3.49)} & {course (5.46)} & {course (3.91)} & {boss (3.06)} & {school (5.16)} \\
          & {4} & {course (3.44)} & {read (3.64)} & {read (3.34)} & {read (2.7)} & {working (2.65)} \\
          & {5} & {class (3.17)} & {class (3.28)} & {team (3.01)} & {working (2.57)} & {team (2.64)} \\    \end{tabular}

    \caption{A comparison of the top-5 words for each LIWC category that appeared in the conversations within each relationship category, along with the proportion of each word.}
    \label{tab:frequent-liwc-words}
\end{table*}

\section*{3. Further details on network metrics}
\label{sec:detail_network}
We denote \(\Gamma(u)\) as the set of neighbors of user \(u\), and \(\mathbf{{m}}_{u\to w}\) as the number of times user \(u\) mentions another user \(w\).
Our metrics for network content are Jaccard coefficient and Adamic-Adar index~\citep{adamic03adamicadar}, both frequently used for measuring the similarity of users in a network.
The Jaccard coefficient measures the percentage of mutual neighbors of the union of neighbors for two individuals as 
\[
    \frac{|\Gamma(u) \cap \Gamma(v)|}{|\Gamma(u) \cup \Gamma(v)|}.
    \]
The Adamic-Adar index also increases with a larger number of mutual neighbors, but is penalized if the mutual neighbor is well-connected. It is defined as
\[\sum_{w \in \Gamma(u) \cap \Gamma(v)} \frac{1}{\log |\Gamma(w)|}.
\]
To allow for direct comparisons among dyads, we use the z-normalized score for each metric instead of the raw score as  
    \[
        zscore(u)=\frac{x-\mu}{\sigma},
    \] 
where \textit{x} is the raw score, \(\mu\) and \(\sigma\) are the mean and standard deviation computed from the neighboring dyads of \textit{u} other than \textit{v}.
For computational efficiency, we sample up to 10 neighbors for computing every metric.

For communication frequency, we measure the following metrics.
We compute the probability of mentioning a specific user out of all possible neighbors, which is obtained as 
    \[\frac{\mathbf{m}_{u\to v}}{\sum_{w \in \Gamma(u)}\mathbf{m}_{u\to w}}.\]
Finally, we compute the reciprocity between two users as the fraction of communications each user has made, denoted as 
    \[
        2\times \frac{\mathbf{min}\left (\mathbf{m}_{u\to v},\mathbf{m}_{v\to u}  \right )}{\mathbf{m}_{u\to v} + \mathbf{m}_{v\to u}}.
    \]
A score of 1.0 means a fully reciprocal dyad with both users communicating equally, and 0 a fully imbalanced dyad where only one mentions the other.

%

\section*{4. Further details on model for relationship prediction}
\label{sec:detail_pred}

\subsection*{4.1. Proposed RoBERTa model}
\label{sec:detail_roberta}
Given a sample dyad containing the tweet interactions and information of the two users, we combine several neural models to obtain vectorized representations.
Each tweet is first tokenized using a pretrained byte-per-encoding (BPE) tokenizer, then is inputted into a RoBERTa base model as a sequence of tokens with length \textit{L}, $\left [ \textup{t}_1,\textup{t}_2,...,\textup{t}_L \right ]$.
The model returns hidden states equal to the number of tokens, $\left [ \mathbf{h}_1,\mathbf{h}_2,...,\mathbf{h}_L \right ]$, 
where each hidden state is a vector of size $\mathbf{h}_t\in \mathbf{R}^d$.
This process is applied to all \textit{N} tweets and retweet interactions within that dyad, resulting in a set of tweet representation vectors, $\left [ \mathbf{t}_1,\mathbf{t}_2,...,\mathbf{t}_N \right ]$
Likewise, we obtain two bio representation vectors $\left [ \mathbf{b}_1,\mathbf{b}_2 \right ]$ by going through the same steps on the bio descriptions from both users.

Usernames are encoded differently, using characters as the units of embedding instead of BPE tokens.
We first created an embedding matrix for the 300 most common characters in all lowercased usernames, then considered each username as a sequence of those characters.
The sequence is transformed into a matrix, which is fed into a series of 1-dimensional convolutional filters~\citep{kim1dcnn}, a widely used method for extracting hidden representations from character-level embeddings.
They are transformed using $\textit{d}_3$, $\textit{d}_4$, $\textit{d}_5$ convolution filters of kernel sizes 3, 4, and 5, then max-pooled and concatenated to result in a name representation vector of size $\mathbf{n}\in \mathbf{R}^d, \textit{d}_3+\textit{d}_4+\textit{d}_5=d$.
Four name representations are obtained with this process: the username and display name for both users in the dyad.

In the first stage of the model we have obtained vector representations for tweets, bio descriptions, and usernames.
The next stage involves merging these features and making actual predictions.
Inspired by the approach of \cite{huang-carley-2019-hierarchical}, we stack the different representation vectors to form a sequence of vectors, $\left [\mathbf{t}_1,\mathbf{t}_2,...,\mathbf{t}_N, \mathbf{b}_1,\mathbf{b}_2,\mathbf{n}_1,\mathbf{n}_2,\mathbf{n}_3,\mathbf{n}_4 \right ]$.
This sequence is fed into a different RoBERTa model, where the goal is to attend to these different representations and obtain hidden representations across different layers of the model.
As the model does not know which vector corresponds to a tweet or a bio, position indices were added to the model.

After six layers of computation within the model, the following hidden states are returned, $\left [\mathbf{o}_1,\mathbf{o}_2,...,\mathbf{o}_{N+6} \right ]$.
We follow common practices in text classification with BERT models and use the first vector $\mathbf{o}_1$ for classification.
The final classification layer composes of two linear transformation layer, a ReLU activation function between the linear transformations, and a softmax function.
For better robustness, dropout~\citep{srivastava14dropout} is applied at a ratio of p=0.1 after ReLU.

Further training details for the task are as follows.
The batch size for both training and testing was 4.
The dimension for the hidden states \textit{d} was set to 768, which is the default value presented in the HuggingFace library.
The initial learning rate was 1e-5, with an initial warmup of 100 steps.
The model was developed in Pytorch~\citep{pytorch} and trained on an NVIDIA GTX 1080Ti graphic card.
The model ran for 5 epochs, with early stopping if the validation score did not improve for 1,000 iterations.
Adam~\citep{kingmaba14adam} with weight decay (eps=1e-8) was used as the optimizer for this model.

\subsection*{4.2. Baseline model}
\label{sec:detail_baseline}
Text features are converted to n-grams using Scikit-learn, where unigrams, bigrams and trigrams with more than 10,000 appearances in the training set were preserved, resulting in 5,377 unique n-grams.
Additionally, 75 and 187 features were generated through lexical count statistics from each dimension of the lexicons, LIWC and Empath~\citep{empath}.
Network features were also added to this model.
The model was trained with XGBoost~\citep{xgboost} on 1,000 rounds with an initial learning rate of 1, with early stopping enabled if the validation loss did not decrease after 20 rounds.

\subsection*{4.3. Diurnal distribution of inferred relationships}

\begin{figure}[t]
  \centering
  \subfigure[Raw frequency, labeled]{\includegraphics[scale=0.18]{figures/cat-500k-1.png}}\quad
  \subfigure[Raw frequency, inferred]{\includegraphics[scale=0.18]{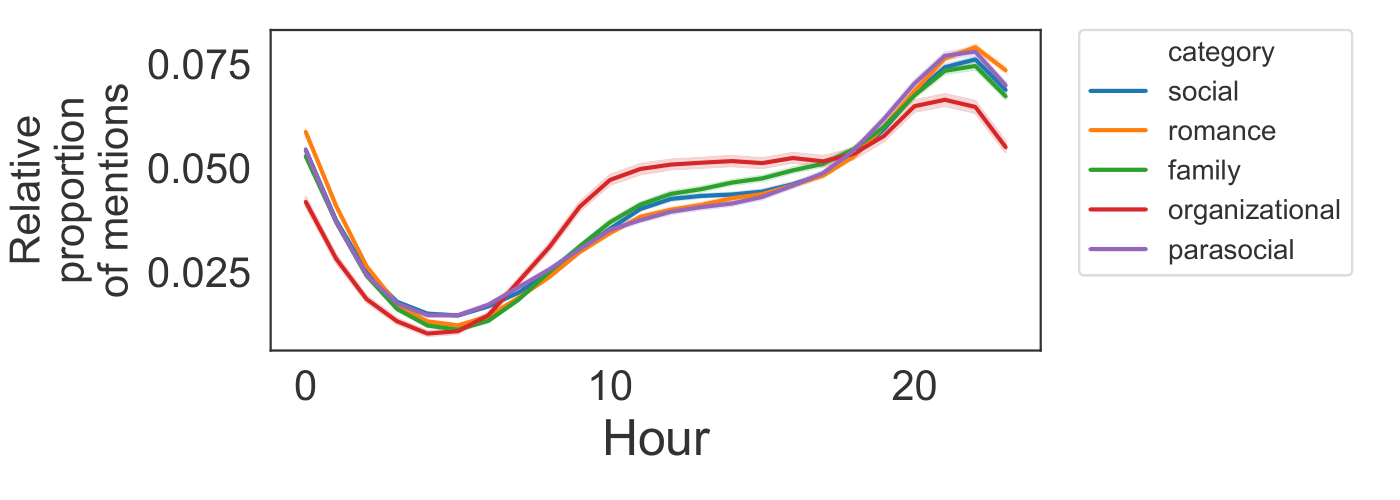}}\quad
  \caption{A comparison of mention frequency across hours of day between dyads with (a) labeled relationships obtained through self-declared mentions, and (b) inferred relationships obtained through the relationship prediction classifiers. Some of the relationship-specific characteristics such as a focus of daytime communication for organization relationships are visible in the inferred categories as well. Shaded regions show 95\% bootstrapped confidence intervals.}
  \label{fig:diurnal_inferred}
\end{figure}

Figure~\ref{fig:diurnal_inferred} shows a comparison between the diurnal distributions calculated from the labeled~(a) and inferred~(b) relationships. We can observe that the inferred dyads share the properties of communicational preference, where again organizational relationships have a higher tendency to commnunicate during the day, compared to other types of relationships.

\section*{5. Further details on model for retweet prediction}
\label{sec:detail_pred_rt}

\subsection*{5.1. Proposed RoBERTa model}
\label{sec:detail_roberta_rt}
Given a tweet posted by one user and a potential retweeting user, we combine information of the tweet with information of the relationship type between the two.

Each tweet is first tokenized using a pretrained byte-per-encoding (BPE) tokenizer, then is inputted into a RoBERTa base model as a sequence of tokens with length \textit{L}, $\left [ \textup{t}_1,\textup{t}_2,...,\textup{t}_L \right ]$.
The model returns hidden states equal to the number of tokens, $\left [ \mathbf{h}_1,\mathbf{h}_2,...,\mathbf{h}_L \right ]$, 
where each hidden state is a vector of size $\mathbf{h}_t\in \mathbf{R}^d$.
We extract the hidden vector representation of this tweet by selecting the first vector, $\mathbf{h}_1$, which is the position of the [CLS] token attached to the beginning of the tweet.

We also obtain two different representations for the relationship between the two users.
First, we create an embedding matrix of size $5 \times d$, where 5 corresponds to the five relationship categories, and \textit{d} is the dimension size of the embedding.
Each \textit{d}-dimensional vector corresponds to the representation of a social, romance, family, organizational and parasocial relationship.
We denote this representation vector as $\mathbf{r}\in \mathbf{R}^d$.
Second, we use the phrase-level information (i.e., my ``best friend'') as well as the relationship category information.
Each character of the phrase is transformed into a vector, resulting into a matrix with a width equal to the number of characters in the phrase.
The resulting matrix is then fed into a series of 1-dimensional convolutional filters~\citep{kim1dcnn}.
They are transformed using $\textit{d}_3$, $\textit{d}_4$, $\textit{d}_5$ convolution filters of kernel sizes 3, 4, and 5, then max-pooled and concatenated, to result in a phrase-level representation vector $\mathbf{p}\in \mathbf{R}^d, \textit{d}_3+\textit{d}_4+\textit{d}_5=d$.

Finally, we combine all available information: representations of (1) the tweet, (2) the relationship category, (3) the phrase for the relationship, and also (4) the number of followers for each users, log-normalized.
The concatenated vector, \[ \mathbf{c}=[\mathbf{h}_1,\mathbf{r},\mathbf{p},fol_u,fol_v], \mathbf{c}\in \mathbf{R}^{3d+2} \]
then goes through (1) linear transformed into a \textit{d}-dimensional vector, (2) a ReLU activation, (3) another linear transformation to a 1-dimensional scalar value, and (4) a sigmoid function that returns a value between 0 and 1, the predicted probability of a retweet happening between the two users.

\clearpage
\section*{6. Additional Performance Details}
\label{sec:full-performance-details}

\subsection*{6.1. Relationship classification performance}
\label{sec:full-rel-details}
Table \ref{tab:all-relationship-prediction-performance} shows the performance of our proposed model for predicting relationships in balanced and imbalanced settings using different subsets of features to quantify the effects of their impact on the overall performance. 
\begin{table*}[t]
    \small
    \centering
    \begin{tabular}{lcccccc}
        & \multicolumn{6}{c}{{\textbf{Performance (F1 score)}}}\\
         {\textbf{Model}}
        & {Social} & {Romance} & {Family} & {Organizational} & {Parasocial} & {Macro F1}\\
         \hline 
         \textbf{{Baselines (balanced)}} \\
         {Random guess} & {0.2} & {0.2} & {0.2} & {0.2} & {0.2} & {0.2} \\
         {GBT Model} & {0.45}& {0.57} & {0.55} & {0.64} & {0.55} & {0.55}  \\
         \textbf{{Proposed model (balanced)}} \\
         {Tweets - directed mentions only} & {0.31} & {0.47} & {0.30} & {0.46} & {0.21} & {0.35}  \\
         {Tweets - public mentions only} & {0.45} & {0.61} & {0.57} & {0.62} & {0.57} & {0.56}  \\
         {Tweets - retweets only} & {0.31} & {0.41} & {0.27} & {0.46} & {0.36} & {0.37}  \\
         {All tweets} & {0.54} & {0.63} & {0.64} & {0.71} & {0.59} & {0.63}  \\
         {All tweets+user profile} & {0.56} & {0.67} & {0.68} & {0.76} & {0.67} & {0.67}  \\
         {All tweets+user profile+network (All features)} & {\textbf{0.60}} & {\textbf{0.69}} & {\textbf{0.69}} & {\textbf{0.79}} & {\textbf{0.72}} & {\textbf{0.70}}  \\
         \hline
         \textbf{{Baselines (Imbalanced)}} \\
         {Random guess} & {0.62} & {0.30} & {0.05} & {0.02} & {0.1} & {0.2}  \\
         {Majority guess} & {0.76} & {0} & {0} & {0} & {0} & {0.15}  \\
         {GBT Model} & {0.80}& {0.52} & {0.33} & {0.28} & {0.28} & {0.44}  \\
         \textbf{{Proposed model (Imbalanced)}} \\
         {All features, trained on balanced data} & {0.72} & {0.69} & {0.38} & {0.35} & {0.29} & {0.49} \\
         {All features, trained on imbalanced data} & {0.84} & {0.68} & {0.51} & {0.50} & {0.38} & {0.58} \\
    \end{tabular}
    
    \caption{Performance comparison on the relationship prediction task for different settings. (1) Public mentions are more informative than directed mentions and retweets. (2) Organizational relationships are easiest to predict across almost all model settings. (3) Even when tested on an imbalanced dataset, our model achieves a decent F1 score of 0.49.}
    \label{tab:all-relationship-prediction-performance}
\end{table*}

\begin{table}[t]
    \centering
    \resizebox{0.48\textwidth}{!}{
    \begin{tabular}{ll cccccc}
        &{\textbf{Category}} & \multicolumn{3}{c}{{{No URL}}} & \multicolumn{3}{c}{{{Has URL}}}\\
         && {Pre.} & {Rec.} & {F-1} & {Pre.} & {Rec.} & {F-1}\\
         \hline 
         &\textit{Overall} & {0.58} & {0.69} & {0.63} & {0.53} & {0.85} & {0.65}\\
         &{Social} & {0.60} & {0.67} & {0.63} & {0.52} & {0.88} & {0.66}\\
        \parbox[t]{2mm}{\multirow{3}{*}{\rotatebox[origin=c]{90}{Baseline}}}
         &{Romance} & {0.60} & {0.69} & {0.64} & {0.53} & {0.92} & {0.67}\\
         &{Family} & {0.58} & {0.66} & {0.62} & {0.55} & {0.85} & {0.66}\\
         &{Organizational} & {0.54} & {0.71} & {0.61} & {0.55} & {0.84} & {0.66}\\
         &{Parasocial} & {0.58} & {0.72} & {0.64} & {0.48} & {0.83} & {0.61}\\

        \hline 
        & \textit{Overall} & {0.58} & {0.71} & {0.64} & {0.53} & {0.87} & {0.66}\\
        \parbox[t]{2mm}{\multirow{3}{*}{\rotatebox[origin=c]{90}{w/ relationship}}}
        & {Social} & {0.59} & {0.72} & {0.65} & {0.53} & {0.94} & {0.68}\\
        & {Romance} & {0.58} & {0.75} & {0.66} & {0.53} & {0.95} & {0.68}\\
        & {Family} & {0.59} & {0.69} & {0.63} & {0.55} & {0.92} & {0.69}\\
        & {Organizational} & {0.54} & {0.65} & {0.61} & {0.55} & {0.86} & {0.66}\\
        & {Parasocial} & {0.58} & {0.72} & {0.64} & {0.50} & {0.79} & {0.61}\\


\end{tabular}
    }
    \caption{Performance metrics for the retweet prediction task that incorporates tweets containing URLs. The scores are grouped into categories of (1) whether the input tweet contains a URL, (2) whether the relationship type was used as an additional feature, and (3) which relationship category the dyad in a sample belongs to. In general, tweets containing URLs are much more likely to be labeled as . For social, romance and family categories, the addition of the relationship type as a feature improves performance through boosting recall.} 
    \label{tab:retweet-prediction-url}
\end{table}

\subsection*{6.2. Retweet prediction performance}
\label{sec:full-rt-details}
Table~\ref{tab:retweet-prediction-url} contains results of the retweet prediction task on settings that do or do not include URL information, trained and tested on each category data.

\section*{7. Further details for the correlation values of the inferred and labeled diurnal distributions in Section 5.4}
Table~\ref{tab:diurnal_coef} displays the Person coefficient values computed between the diurnal distributions between the labeled and inferred relationships.
\begin{table}[]
    \centering
    \begin{tabular}{llcc}
        Inferred & Labeled & correlation coef. & p-val \\
        \hline
	family &	family &	0.949 &	0.000 \\
	family &	organizational &	0.749 &	0.000 \\
	family &	parasocial &	0.970 &	0.000 \\
	family &	romance &	0.971 &	0.000 \\
	organizational &	family &	0.988 &	0.000 \\
	organizational &	organizational &	0.928 &	0.000 \\
	organizational &	parasocial &	0.969 &	0.000 \\
	organizational &	romance &	0.975 &	0.000 \\
	parasocial &	family &	0.920 &	0.000 \\
	parasocial &	organizational &	0.677 &	0.000 \\
	parasocial &	parasocial &	0.947 &	0.000 \\
	parasocial &	romance &	0.954 &	0.000 \\
	romance &	family &	0.899 &	0.000 \\
	romance &	organizational &	0.653 &	0.001 \\
	romance &	parasocial &	0.930 &	0.000 \\
	romance &	romance &	0.935 &	0.000 \\
	social &	family &	0.927 &	0.000 \\
	social &	organizational &	0.699 & 	0.000 \\
	social &	parasocial &	0.952 &	0.000 \\
	social &	romance &	0.957 &	0.000 \\
        
    \end{tabular}
    \caption{The Pearson coefficients computed between the diurnal distributions of labeled and inferred relationship categories.}
    \label{tab:diurnal_coef}
\end{table}

\section*{8. Further details for reproducibility}
Table~\ref{tab:reproducibility} contains details of additional information in the experiments, to better ensure future reproducibility.
\begin{table*}[t]
    \centering
    \begin{tabular}{p{6cm}p{9cm}}
        Category & Description \\
        \hline
        Description of computing infrastructure used & GTX 1080Ti (GPU), Ubuntu 16.04 (OS) \\
        Bounds for hyperparameter search (relationship classification) & lr=[1e-3, 3e-4, 1e-4, 3e-5, 1e-5, 3e-6, 1e-6] \\
        Bounds for hyperparameter search (retweet prediction) & lr=[1e-3, 3e-4, 1e-4, 3e-5, 1e-5, 3e-6, 1e-6] \\
        Criterion for hyperparameter search in relationship classification & macro F-1 score on validation set \\
        Criterion for hyperparameter search in retweet prediction & AUC score on validation set \\
        Average runtime for relationship prediction & 10hrs per epoch, 5 epochs (balanced setting)\newline20hrs per epoch, 5 epochs (imbalanced setting) \\
        Average runtime for retweet prediction & 2hrs per epoch, 10 epochs (balanced setting)\newline9hrs per epoch, 5 epochs (imbalanced setting) \\
        Number of parameters in relationship classification model & 148,150,253 (proposed model) \\
        Number of parameters in retweet prediction model & 136,479,748 (baseline model)\newline137,417,656 (proposed model) \\
        Validation score of best-performing relationship classification model & F-1: 0.672 (proposed, balanced set)\newline F-1: 0.559 (proposed, imbalanced set) \\
        Validation score of best-performing retweet prediction model & AUC: 0.633 (baseline)\newline AUC: 0.640 (proposed) \\

    \end{tabular}
    \caption{Description of criteria for reproducibility}
    \label{tab:reproducibility}
\end{table*}
\end{appendices}

\end{document}